\documentstyle[epsf,11pt,twoside]{IEEEtran}

\begin{document}

\bibliographystyle{IEEEtran}


\newcommand{\figureSingleN}[1]{
  \begin{figure}[t]{\sloppy \footnotesize#1}\end{figure}
}

\newcommand{\figureSingleW}[1]{
  \begin{figure*}[tb]{\sloppy \footnotesize#1}\end{figure*}
}

\newcommand{\figurePair}[2]{
  \begin{figure*}[t]
    \begin{minipage}[t]{0.48 \textwidth}{\sloppy \footnotesize#1}\end{minipage}
    \hfill
    \begin{minipage}[t]{0.48 \textwidth}{\sloppy \footnotesize#2}\end{minipage}
  \end{figure*}
}

\def\keywords{\vspace{-.3em}
    \if@twocolumn
      \small\it Keywords\/\bf---$\!$%
    \else
      \begin{center}\small\bf Keywords\end{center}\quotation\small
    \fi}
\def\endkeywords{\vspace{0.6em}\par\if@twocolumn\else\endquotation\fi
    \normalsize\rm}

\newenvironment{biographyPIC}[2]{%
\footnotesize\unitlength 1mm\bigskip\bigskip\bigskip\parskip=0pt\par%
\rule{0pt}{39mm}\vspace{-39mm}\par
\noindent\setbox0\hbox{\epsfxsize=25mm \epsfysize=32mm \epsfbox{#2}}%
\ht0=37mm\count10=\ht0\divide\count10 by\baselineskip
\global\hangindent29mm\global\hangafter-\count10%
\hskip-28.5mm\setbox0\hbox to 28.5mm {\raise-30.5mm\box0\hss}%
\dp0=0mm\ht0=0mm\box0\noindent\bf#1\rm}{
\par\rm\normalsize}


\setcounter{page}{101}

\markboth{IEEE Transactions On Visualization and Computer Graphics,
Vol. 4, No. 4, December 1998}{Hodgins et al.: Perception of Human~Motion
with Different Geometric~Models}


\renewcommand{\thefootnote}{\fnsymbol{footnote}}

\title{Perception of Human~Motion with Different Geometric~Models}
\author{Jessica K. Hodgins, James F. O'Brien, Jack Tumblin$^{\dag}$}
\maketitle

\footnotetext[2]{\fussy\it
  The authors are with the Graphics, Visualization, and Usability
  Center, College of Computing, Georgia Institute of Technology,
  Atlanta, GA 30332.  E-mail: \{jkh,obrienj,ccsupjt\}@cc.gatech.edu
}
\renewcommand{\thefootnote}{\arabic{footnote}}
\pretolerance=900\tolerance=1200
\sloppy


\begin{abstract}
  Human figures have been animated using a variety of geometric models
  including stick figures, polygonal models, and NURBS-based models
  with muscles, flexible skin, or clothing.  This paper reports on
  experimental results indicating that a viewer's perception of motion
  characteristics is affected by the geometric model used for
  rendering.  Subjects were shown a series of paired motion sequences
  and asked if the two motions in each pair were {\it the same} or
  {\it different}.  The motion sequences in each pair were rendered
  using the same geometric model.  For the three types of motion
  variation tested, sensitivity scores indicate that subjects were
  better able to observe changes with the polygonal model than they
  were with the stick figure model.
\end{abstract}

\begin{keywords}
  Motion perception, motion sensitivity, computer animation, geometric
  model, perceptual study, biological motion stimuli, light-dot display.
\end{keywords}


\section{Introduction} \label{sec-intro}
Few movements are as familiar and recognizable as human walking and
running.  Almost any collection of dots, lines, or shapes attached to
an unseen walking figure is quickly identified and understood as
human.  Studies in human perception have displayed walking motion
using only dots of light located at the joints and have found test
subjects quite adept at assessing the nature of the underlying
motion\cite{Johansson:1973:vpbmma}.  In particular, subjects can
identify the gender of a walker and recognize specific individuals
from light-dot displays even when no other cues are
available\cite{Cutting:1977:rftw,Kozlowski:1977:rswdpd,Kozlowski:1978:rgwplmasst}.

In part because people are skilled at detecting subtleties in human
motion, the animation of human figures has long been regarded as an
important, but difficult, problem in computer animation.  Recent
publications have presented a variety of techniques for creating
animations of human motion.  Promising approaches include techniques
for manipulating keyframed or motion capture
data\cite{Bruderlin:1995:MSP,Rose:1996:EGMT,Unuma:1995:FPE,Witkin:1995:MW},
control systems for dynamic
simulations\cite{Hodgins:1995:AHA,Laszlo:1996:LCCAABW,Ngo:1993:SCR,Panne:1995:GOBL,Panne:1993:SAN},
and other procedural or hybrid
approaches\cite{Badler:1993:SHC,Bruderlin:1989:GDD,Cohen:1992:ISC,Girard:1985:CMC,Ko:1993:SLW,Laurent:1992:it,Perlin:1995:RTR}.
Each method has its own strengths and weaknesses, making the visual
comparison of results essential, especially for the evaluation of such
subjective qualities as naturalness and emotional expression.  
The research community has not yet adopted a standard set of models and 
there is currently enormous variety in the models and rendering styles 
used to present results.

Our ability to make judgments about human motion from displays as
rudimentary as dot patterns raises an important question: Does the
geometric model used to render an animation affect a viewer's judgment
of the motion or can a viewer make accurate judgments independent of
the geometric model?  There are three plausible but contradictory
answers to this question.

\vskip 0.0625in
{\bf Possibility 1. Simple representations may allow finer
distinctions when judging human motion.}
Simpler models may be easier to comprehend than more complex ones,
allowing the viewer's attention to focus more completely on the
details of the movement rather than on the details of the model.  For
example, a stick figure is an obvious abstraction and rendering flaws
may be easily ignored.  When more detailed models are used, subtle
flaws in rendering, body shape, posture, or expression may draw
attention away from the movements themselves.  Complex models may also
obscure the motion. For example, the movement of a jacket sleeve might
hide subtle changes to the motion of the arm underneath.

\vskip 0.0625in
{\bf Possibility 2. Complex, accurate representations may allow finer
distinctions.}
People have far more experience judging the position and movement of
actual human shapes than they do judging abstract representations
such as stick figures.  A viewer, therefore, may be able to make finer
distinctions when assessing the motion of more human-like
representations.  Furthermore, complex representations provide more
features to identify and track.  Each body segment in a polygonal
human model has a distinctive, familiar shape, making it
easier to gauge fine variations in both position and rotation.

\vskip 0.0625in
{\bf Possibility 3. Both simple and complex representations may allow
equally fine distinctions.}
The human visual system may use a displayed image only to maintain the
positions of a three-dimensional mental representation.  Judgments
about the motion may be made from this mental representation rather
than directly from the viewed image.  Displayed images must of course
supply enough cues to keep the mental representation accurate, but
additional detail and accuracy may be irrelevant.  Just as joint
positions shown by light dots are sufficient to control the mental
representation, connecting the dots with a stick figure might not
improve the viewer's perception.  Similarly, encasing a stick figure
within a detailed human body shape might likewise prove unnecessary.
\vskip .2in

Objective evidence is needed to determine which of these possibilities
is correct.  We argue that definitive experiments to select between
possibilities~1 and~2 are impractical.  The question of which style of
geometric model is more useful for judging motion is likely to be
highly complex and context dependent, affected by all of the variables
of both the motion and the rendering.  If possibility~3 were correct
and model style were largely irrelevant, then we would be able to
perform critical comparisons of the motion synthesis techniques in the
literature by direct comparison of the substantially different
geometric models used in each publication.  This paper provides
experimental evidence to disprove possibility~3 by showing that viewer
sensitivities to variations in motion are significantly different for
the stick figure model and the polygonal model shown in
Fig.~\ref{fig:polyAndStick}.  In particular, for the types of motion
variation we tested, viewers were more sensitive to motion changes
displayed through the polygonal model than through the stick figure
model.  This result suggests that stick figures may not always have
the required complexity to ensure that the subtleties of the motion
are apparent to the viewer.

\figureSingleW{
  \centerline{\epsfysize=2.25in \epsfbox{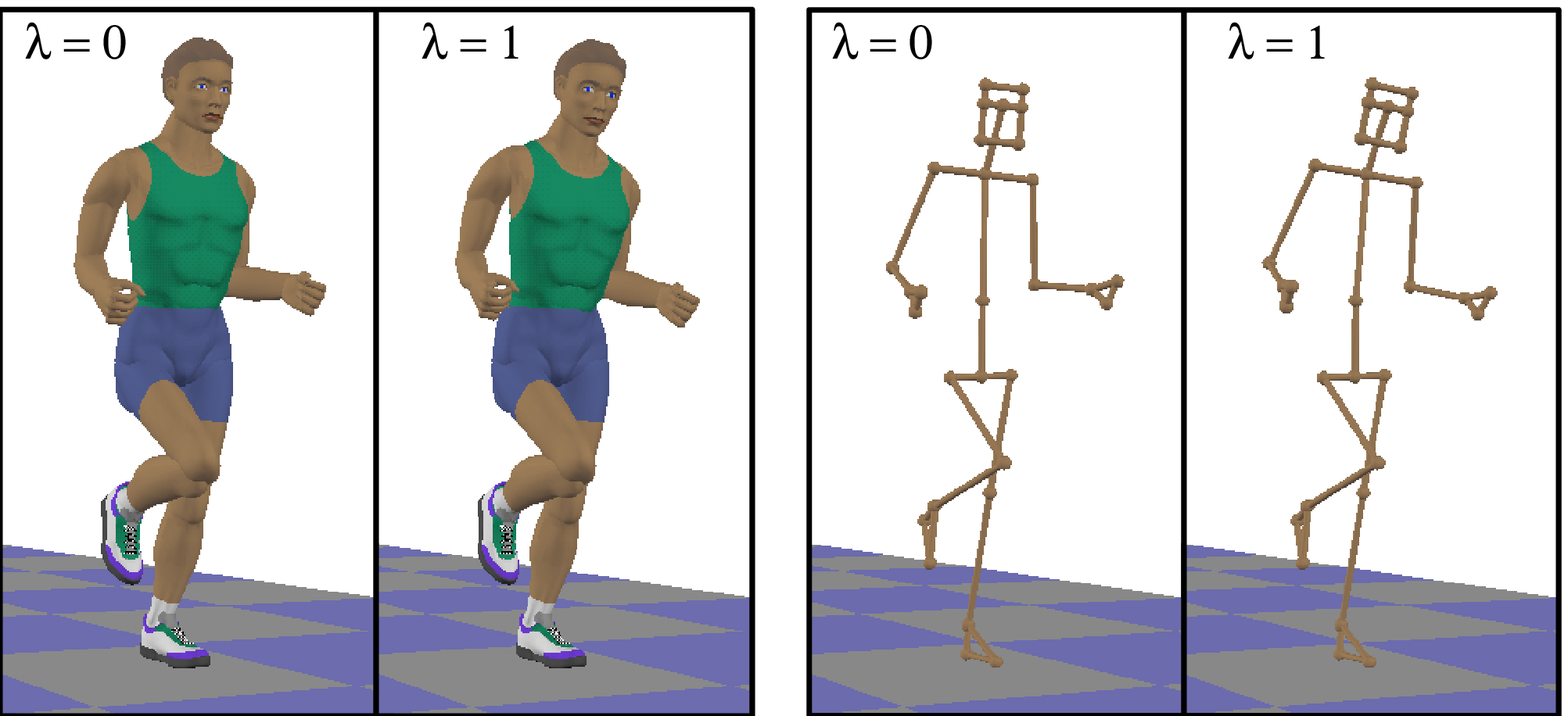}}
  \caption{
    Images of an animated human runner.  The pair on the left compares
    two running motions rendered using a polygonal model.  On the
    right, the same pair of motions are rendered with a stick figure
    model.  Modifications to the motion were controlled by a
    normalized parameter,~$\lambda$, that varied between $\lambda=0$
    and $\lambda=1$.  These images are from the motion generated for
    the additive noise test discussed in Section~\ref{subsec-noise}.
    The difference in posture created by the additive noise can be
    seen in the increased angle of the neck and waist in the right
    image of each pair ($\lambda=1$).  
  } \label{fig:polyAndStick}
}


\section{Background} \label{sec-background}

Several researchers have used light-dot displays, also referred to as
biological motion stimuli, to study perception of human movements and
to investigate the possibility of dynamic mental
models\cite{Freyd:1987:dmr}.  The light-dot displays show only dots or
patches of light that move with the main joints of walking figures
(Fig.~\ref{fig:dots}), but even these minimal cues have been shown
to be sufficient for viewers to make detailed assessments of the
nature of both the motion and the underlying figure.  

\figureSingleN{
  \centerline{\epsfxsize=3in \epsfbox{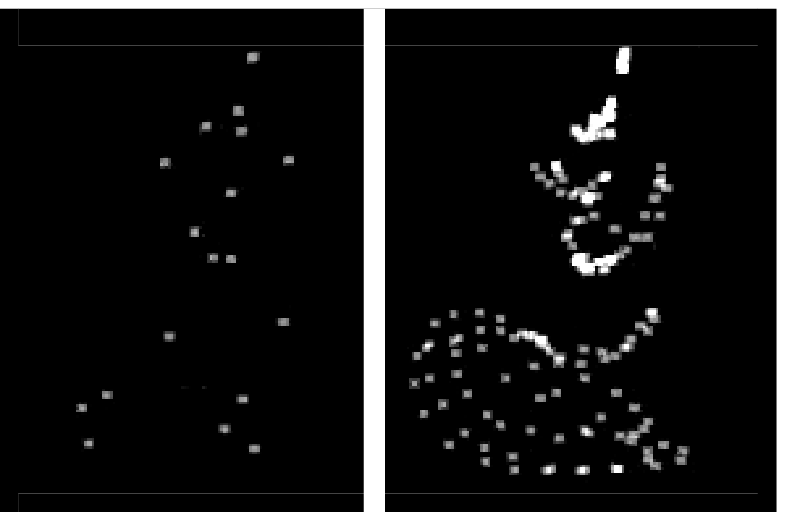}}
  \caption{
    The dot pattern on the left shows the joint locations of a human
    runner at a single point in time.  On the right, these joint
    locations are shown over the course of one step in the running
    cycle.  Although it is difficult to determine the nature of these
    patterns from a still image, studies show that most people are
    able to recognize the motion and even to make fine judgments when
    shown moving sequences of similar images.
  } \label{fig:dots}
}
 
The ability to perceive human gaits from light-dot displays has been
widely reported to be acute and robust.  Early experiments by
Johansson reported that 10-12~light dots ``evoke a compelling
impression of human walking, running, dancing,
etc.''\cite{Johansson:1973:vpbmma}.  Because such displays provide
motion cues independent of form or outline, other investigators have
used them to study human motion perception.  Work by Cutting and
Kozlowski showed that viewers easily recognized friends by their
walking gaits on light-dot displays\cite{Cutting:1977:rftw}.  They
also reported that the gender of unfamiliar walkers was readily
identifiable, even after the number of lights had been reduced to just
two located on the ankles\cite{Kozlowski:1977:rswdpd}.  In a published
note, they later explained that the two light-dot decisions were
probably attributable to stride
length\cite{Kozlowski:1978:rgwplmasst}.  Continuing this work,
Barclay, Cutting, and Kozlowski showed that gender recognition based
on walking gait required between 1.6~and 2.7~seconds of display, or
about two step cycles\cite{Barclay:1978:tasfigptigr}.  Our experiments
used pairs of running stimuli 4~seconds in duration that displayed
about six strides.  We noticed that test subjects often marked
their answer sheets near the midpoint of the second stimuli which is
consistent with Barclay's results.

Motion is apparently essential for identifying human figures on
light-dot displays.  The Cutting studies reported that while moving
light-dot displays were recognized immediately, still light-dot
displays of a walking figure were not recognized as human.  Poizner
and colleagues also noted that movement is required for accurately
reading American Sign Language gestures\cite{Poizner:1981:poaslidpld}.

This capacity to recognize moving figures was shown to be robust in
the presence of masking by additional light points.  In a modified
experiment, subjects were shown light-dot displays of walkers facing
either left or right and asked to determine walking direction.  Only
complex masks of extraneous light dots moving in patterns that were
similar to those of the walking figure were able to disrupt viewer
judgments\cite{Cutting:1988:mtmohg}.

Appropriate synthetic movements are easily accepted as human when
rendered as light-dot displays.  Cutting and colleagues
found that apparent torso structure and rotation were strongly
correlated with judgments of walker gender\cite{Cutting:1978:abifgp}.
Cutting then constructed a simple mathematical model of light-dot
motion for human walkers and computed displays of synthetic walkers.
Viewers easily identified the synthetic displays as human walkers and
accurately determined the intended gender of the walkers.  These
experiments clearly showed that variations in torso rotation are
important for gender judgments.  Accordingly, we chose to measure
viewer sensitivity to torso rotations in one of our experiments.
 
Proffitt and colleagues found that occlusion of light dots by clothing
or human body segments plays an important role in gait judgment and
may also provide information about body
outlines\cite{Proffitt:1978:trooirmimpld}.  Synthetic displays without
occlusion yielded poorer subject performance.  These experimental
observations suggest that extremely simple models of human figures,
such as thin stick figures, may present similar difficulties for
the~viewer.

Surprisingly, the perception of rigid body segments between moving
light dots at joints does not generalize to movements of isolated
pairs of light dots.  Ishiguchi showed test subjects one fixed light
dot and a second one that moved on an arc of $\pm15$ degrees as if it
were on the end of a pendulum with the first light dot as the pivot
joint\cite{Ishiguchi:1988:teoooiesfdm}.  Viewers perceived the dots as
attached to a flexible bar held fixed at the first light dot rather
than as a rigid bar moving as a pendulum.  Thus the perception of
rigid body segments in the largely pendulum-like movements of human
walking is exceptional; perhaps the ensemble of dots is important, or
perhaps the movements are so intimately familiar that the perception
of an assembly of flexible bars is overridden.


\section {Experimental Methods}

\figureSingleW{
  \centerline{\epsfxsize=140.0mm \epsfbox{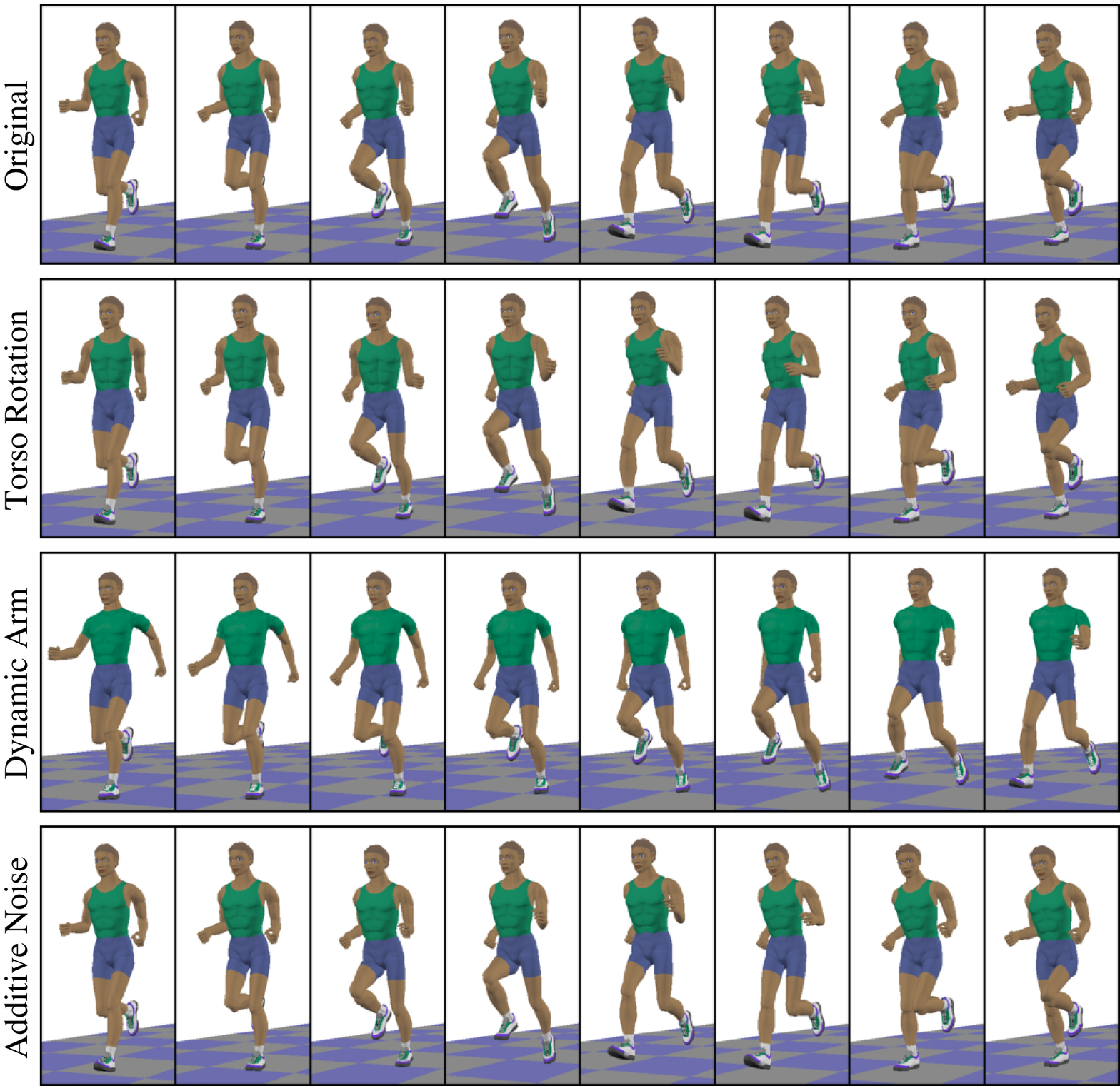}}
  \caption{
    Examples from the motion sequences rendered with the polygonal model.
    {\bf First Row:} Original motion sequence, $\lambda=0$, used in all tests. 
    {\bf Second Row:} Torso rotation motion sequence with $10\times$
      magnification of the torso rotation, $\lambda=1$.
    {\bf Third Row:} Dynamic arm motion sequence with maximum exaggeration,
      $\lambda=1$.
    {\bf Fourth Row:} Additive noise motion sequence with sinusoidal noise of
      $\pm0.15$~radians, $\lambda=1$.  
    Images are spaced at intervals of $0.067$ seconds.
  }\label{fig:motPoly}
}

\figureSingleW{
  \centerline{\epsfxsize=140.0mm \epsfbox{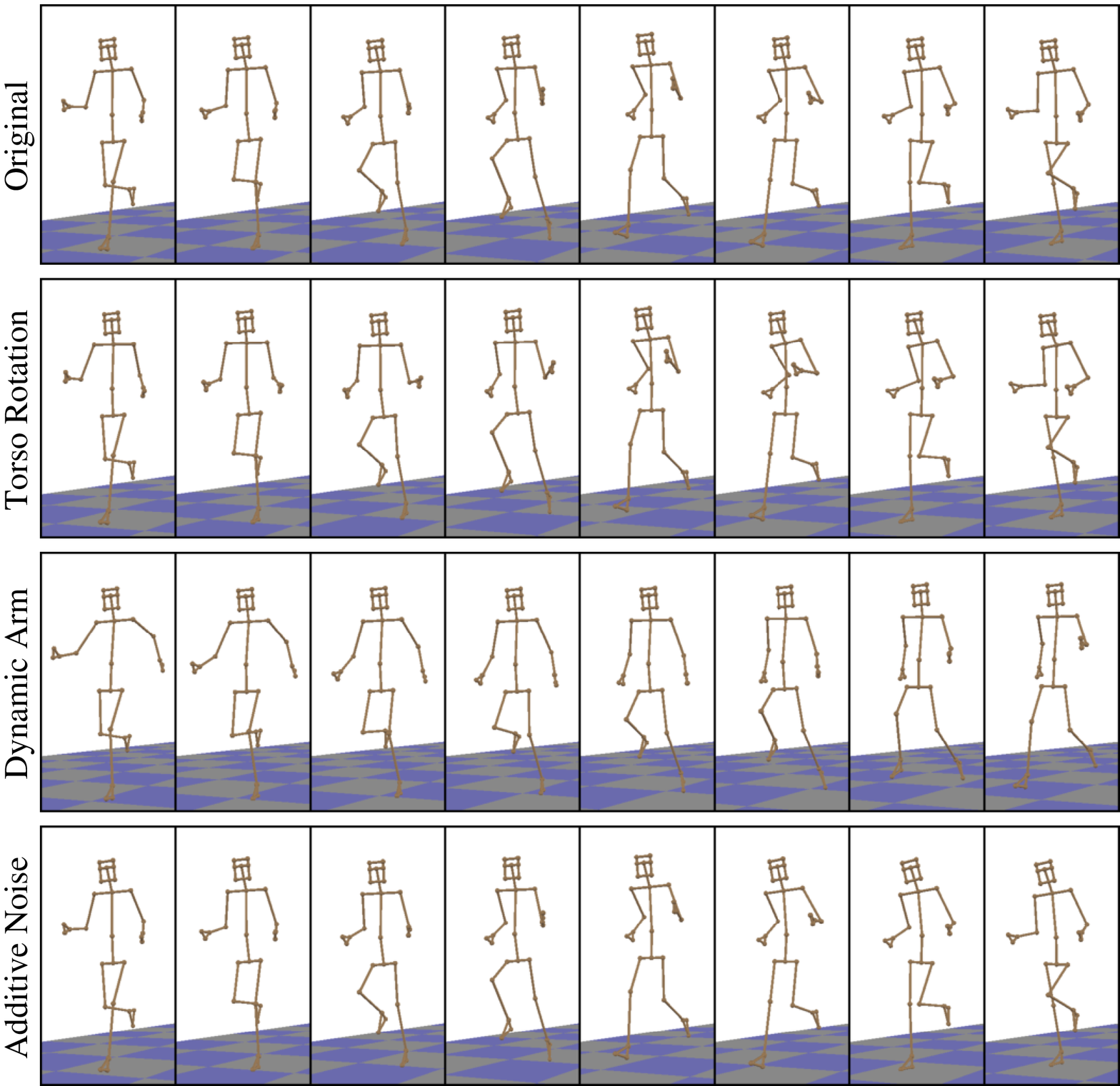}}
  \caption{
    Examples from the motion sequences rendered with the stick figure
      model.
    {\bf First Row:} Original motion sequence, $\lambda=0$, used in all tests.
    {\bf Second Row:} Torso rotation motion sequence with $10\times$
      magnification of the torso rotation, $\lambda=1$.
    {\bf Third Row:} Dynamic arm motion sequence with maximum exaggeration,
      $\lambda=1$.
    {\bf Fourth Row:} Additive noise motion sequence with sinusoidal noise of
      $\pm0.15$~radians, $\lambda=1$.
    Images are spaced at intervals of $0.067$ seconds.
  } \label{fig:motStik}
}

While it is impossible to exhaustively test all of the variables that
may affect a perceived motion, we can use A/B comparison tests to form
a preliminary assessment of whether the geometric model affects a
viewer's perception of motion.  We evaluated three different types of
motion variation in separate experiments described below: torso
rotation, dynamic arm motion, and additive noise.  For each
experiment, the modifications to the motion were controlled by a
normalized parameter, $\lambda$, that varied between $\lambda=0$ and
$\lambda=1$.  \mbox{Figs.~\ref{fig:motPoly}}
\mbox{and~\ref{fig:motStik}} show sequences of images excerpted
from the base motion sequence, $\lambda=0$, and the modified
sequences, $\lambda=1$, used in all three experiments.  Joint angle
trajectories are shown in Fig.~\ref{fig:dplts} to illustrate the key
components of the base motion and the modified motions created by
setting $\lambda=1$.

In each of the three tests, subjects viewed pairs of animated
sequences rendered using the same geometric model and were asked
whether the motions in the two sequences were the same or different.
We then computed a sensitivity measure for each type of geometric
model.  The difference between the sensitivity values is a measure of
whether a particular subject was better able to discriminate between
the motions when they were rendered with a polygonal model or with a
stick figure model.


\subsection {Experiment One: Torso Rotation}

This experiment measured whether a subject's ability to differentiate
between larger and smaller yaw rotations of a runner's torso was
affected by the geometric model used for rendering.  The motion
sequences were generated by making kinematic modifications to data
obtained from a physically based dynamic simulation of a human
runner\cite{Hodgins:1995:AHA}.  The torso's rotation about the
longitudinal axis, or yaw relative to the pelvis, was exaggerated
(Fig.~\ref{fig:showRot}--A).  The neck was counter-rotated to
compensate for the torso rotation so that the facing direction of the
head remained unchanged.

\figureSingleN{
  \centerline{
    \epsfysize=2in \epsfbox{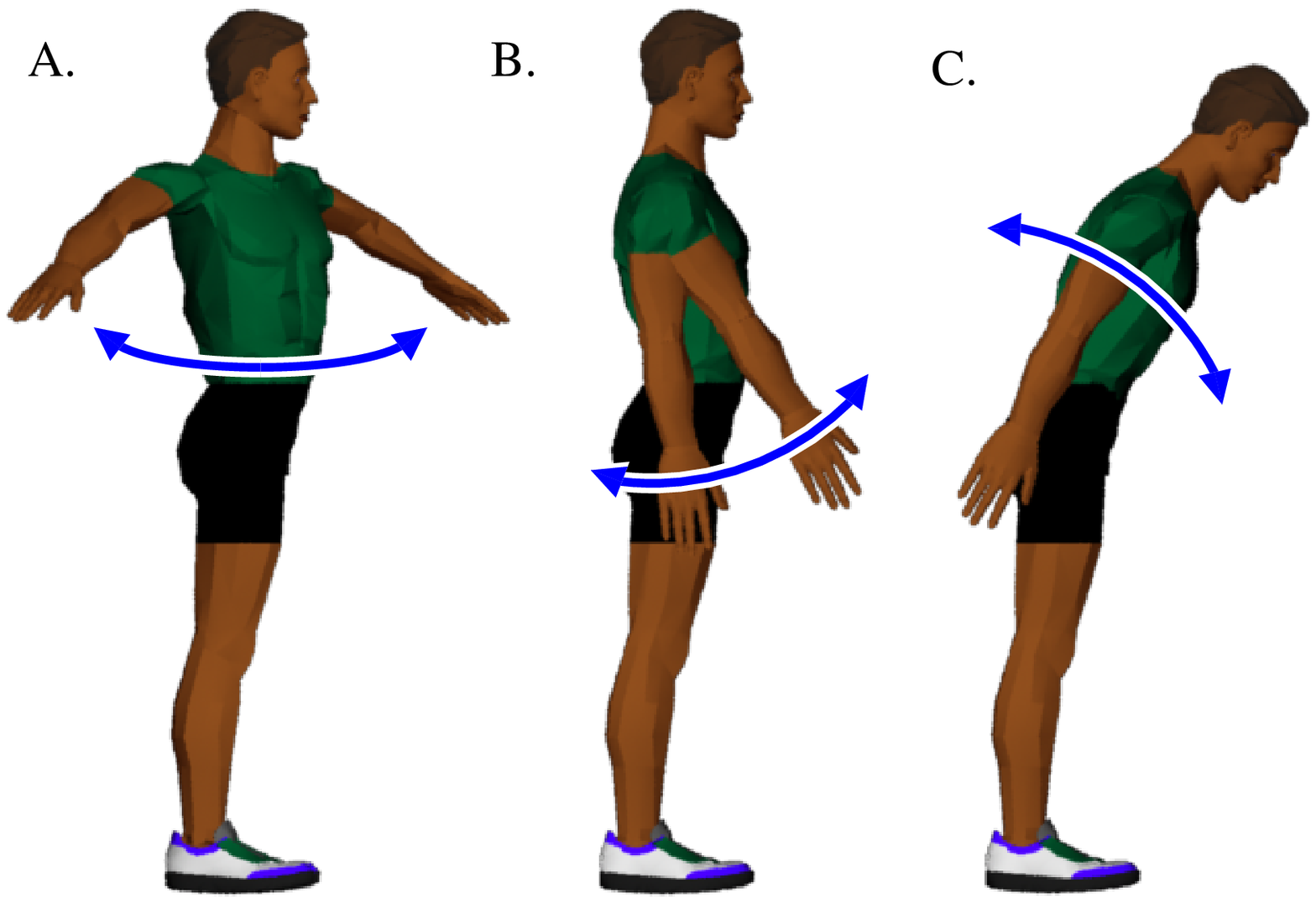} 
  }
  \caption{
    Degrees of freedom for data plotted in Fig.~\ref{fig:dplts}.
    {\bf A.}~Rotation of torso at waist about longitudinal axis relative to pelvis.
    {\bf B.}~Rotation of arm at shoulder about transverse axis relative to torso.
    {\bf C.}~Rotation of torso at waist about transverse axis relative to pelvis.
  } \label{fig:showRot} 
  \vskip 0.05in
}

\figureSingleW{
  \centerline{
    \epsfxsize=2in \epsfbox{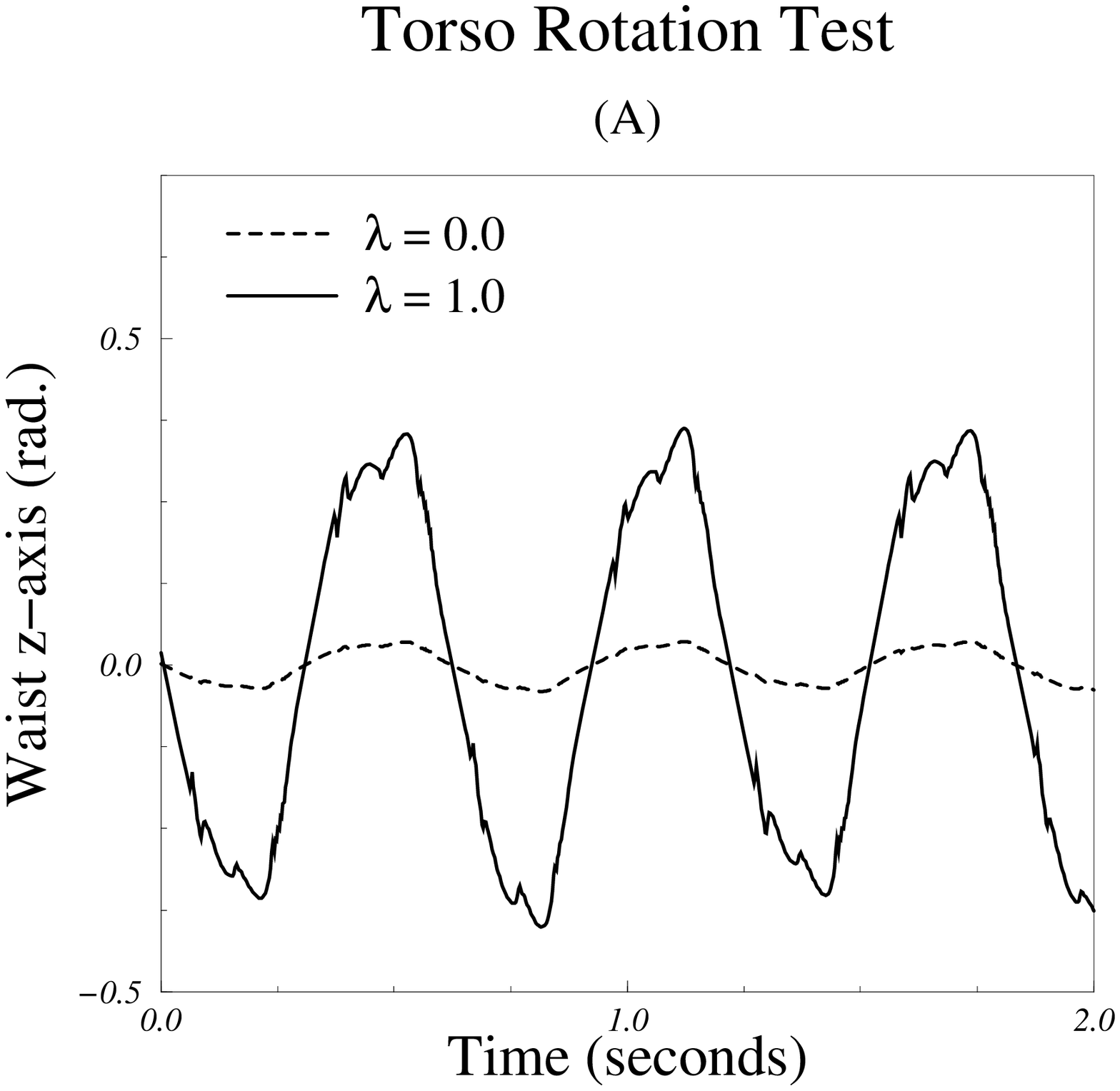} 
    \epsfxsize=2in \epsfbox{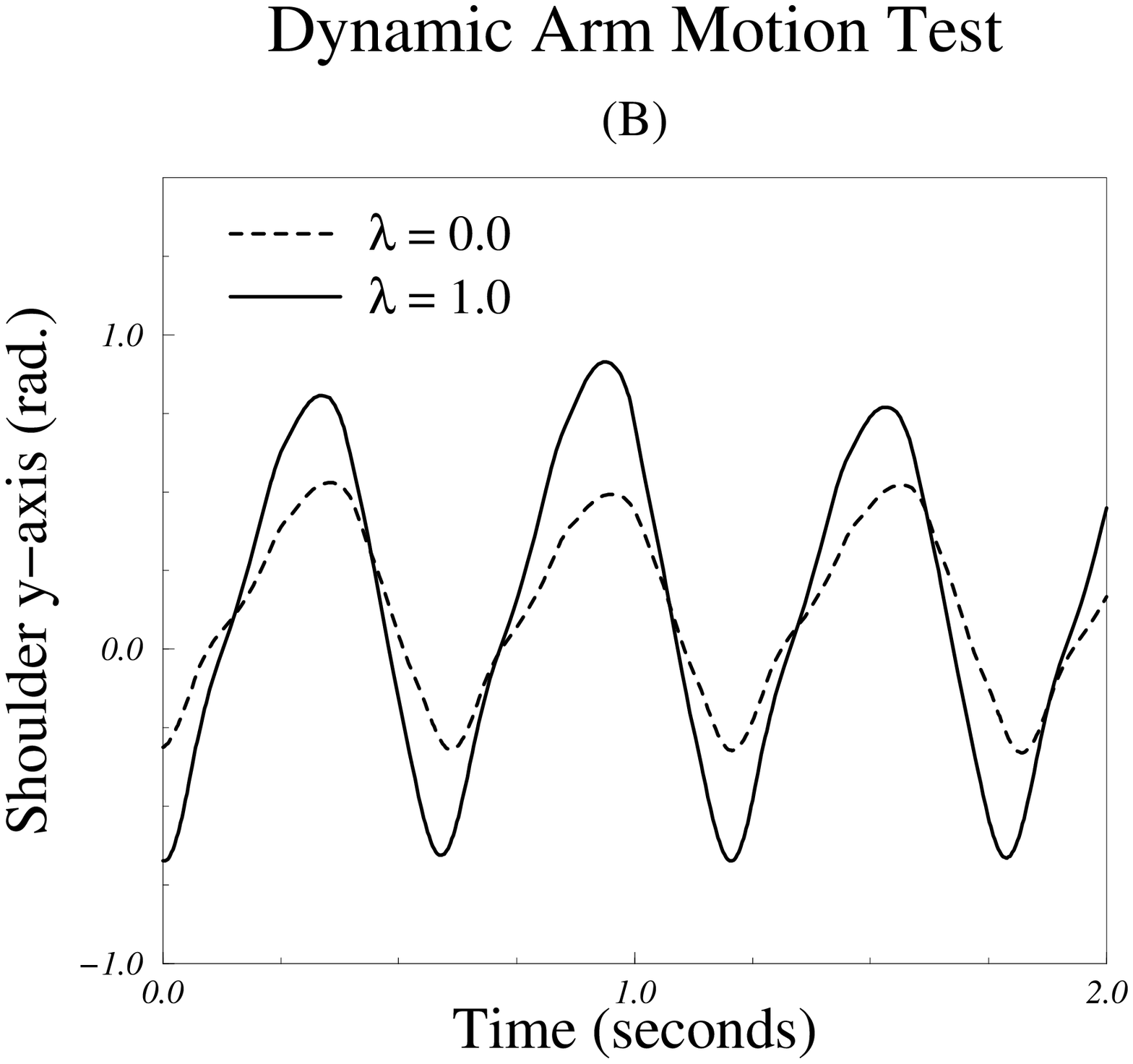} 
    \epsfxsize=2in \epsfbox{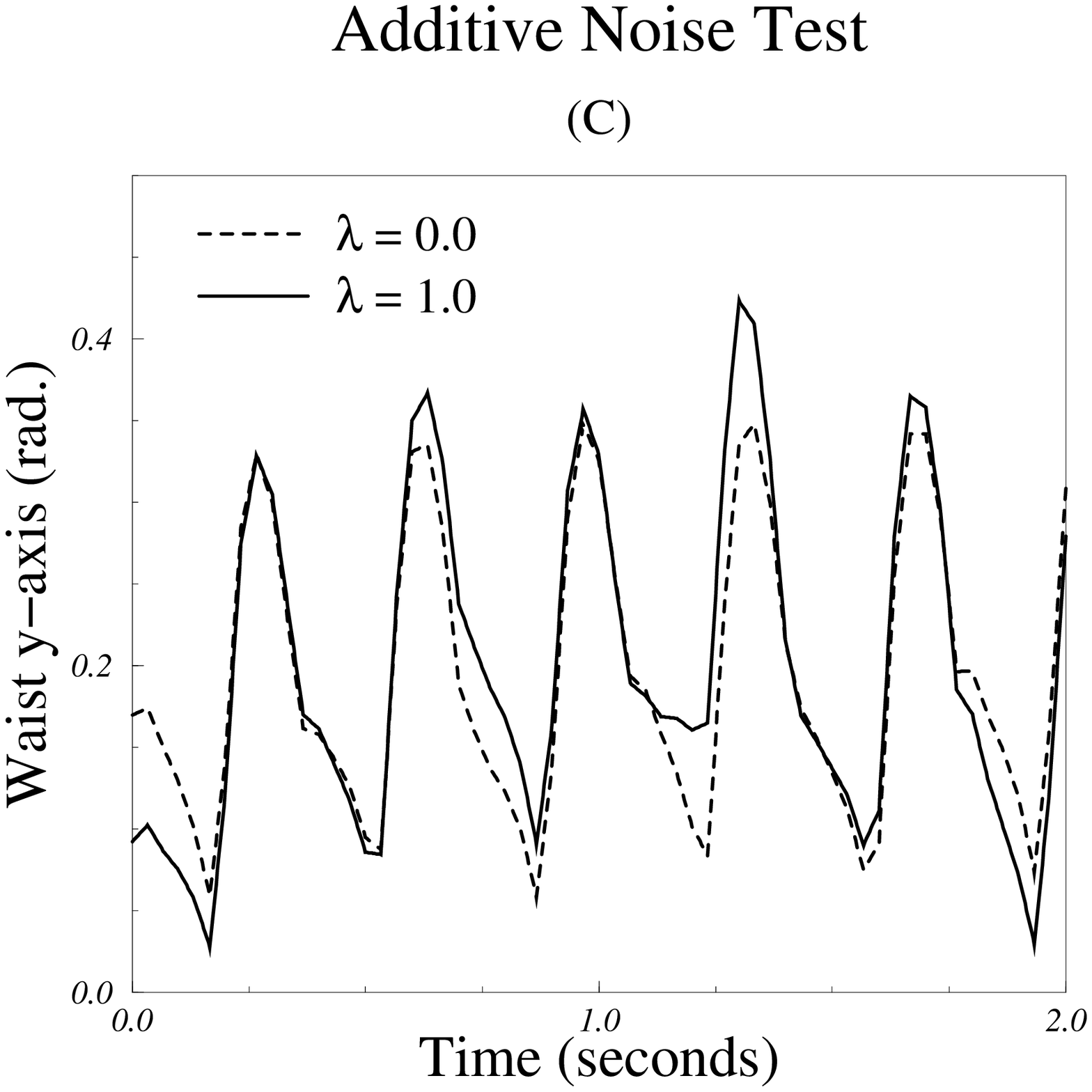}
  }
  \caption{
    Selected joint angle trajectories demonstrating motion differences
      plotted for base motion, $\lambda = 0$, and for extremes of
      modified motion, $\lambda = 1$.
    {\bf A.}~Rotation of torso at waist about the longitudinal axis
      (\textit{z}--axis) for torso rotation test.
    {\bf B.}~Shoulder angle about the transverse axis
      (\textit{y}--axis) for dynamic arm motion test.
    {\bf C.}~One representative modification for the additive noise
      test: the rotation of torso at waist about the transverse axis
      (\textit{y}--axis).
  } \label{fig:dplts} 
}

The magnitude of the exaggeration in torso rotation was controlled by
a normalized parameter, $\lambda$.  A value of $\lambda=0$ gave a
magnification factor of $1\times$ so that the modified motion was
identical to that of the original data.  Larger values of $\lambda$
correspond linearly to higher magnification factors, with $\lambda =
1$ yielding a $10\times$ magnification of the torso rotation. The
motion of body segments below the waist was left unchanged
(Figs.~\ref{fig:motPoly},~\ref{fig:motStik},
and~\ref{fig:dplts}--A).

The test consisted of a series of 40~pairs of motion sequences divided
into two sets of 20~pairs each.  One set was rendered with the stick
figure model and the other with the polygonal model
(Fig.~\ref{fig:polyAndStick}).  All other parameters used to render
the animations, such as lighting, ground models, and camera motion,
were identical for the two sets.  Within each set, half of the pairs
were randomly selected to show two different motion sequences
(different $\lambda$ values).  Of those that were different, the pairs
with the largest disparity in $\lambda$ were placed toward the
beginning of each set so that the questions became progressively more
difficult.  To minimize bias due to fatigue or learning effects, we
varied the order in which the two sets were presented.  Asymmetric
learning effects would not necessarily be minimized by this ordering.

Twenty-six student volunteers who were not familiar with the animations
served as subjects.  All had normal or
corrected-to-normal vision.  Subjects were tested in groups of two or
three in a quiet room.  They were instructed to remain silent and not to collaborate during the test.
The test stimulus was presented on a 20-inch
monitor approximately three feet from the subjects.  All animations
were prerendered and shown at 30~frames per second in NTSC
resolution.  These experimental conditions were selected because they
match the viewing conditions commonly encountered when watching
animated motion.

Subjects were told that they would be shown a series of 4-second
computer-generated animations of a human runner and that the
animations would be grouped in A/B~pairs with 5~seconds of delay
between the presentation of each pair.  Subjects were asked to view
each pair and then indicate on a response sheet whether the two
motions were the same or different.  They were also informed that the
variations would be confined to the motion of the runner's upper body
and that the questions would become progressively more difficult.  A
monetary reward for the highest percentage of correct responses was
offered as an incentive to all test subjects.  Subjects were not told
the purpose of the experiment.


\subsection {Experiment Two: Dynamic Arm Motion}

This experiment measured whether a subject's ability to differentiate
between larger and smaller arm motions was affected by the geometric
model used for rendering.  The motion sequences were generated by
modifying the desired fore-aft rotation about the transverse axis at
the shoulder joint in the dynamic simulation of the human runner
(Fig.~\ref{fig:showRot}--B).  The control routines then computed
torques based on the desired value of the shoulder joint.  These
torques were applied to the dynamic model. The resulting motion is
shown in Figs.~\ref{fig:motPoly},~\ref{fig:motStik},
\mbox{and~\ref{fig:dplts}--B}.  Because the motion was dynamically
simulated, the exaggerated arm motion also had subtle effects on other
aspects of the running motion.

The magnitude of the exaggeration in arm motion was controlled by a
normalized parameter,~$\lambda$.  A value of $\lambda = 0$ gave a
magnification factor of $1\times$ so that the modified motion was
identical to that of the original data.  Larger values of $\lambda$
correspond linearly to higher magnification factors, with $\lambda =1$
yielding a $1.5\times$ magnification of the shoulder rotation.

Twenty-four student volunteers who had not participated in the first
experiment were subjects for this second experiment.  Testing
procedures and format were identical to those used in the first
experiment.


\subsection {Experiment Three: Additive Noise} \label{subsec-noise}

The format of this experiment was identical to that of the first two,
except for the manner in which the running motion was modified.  For
this experiment, time-varying noise was added to the joint angles for
the waist, shoulders, and neck.  The noise was generated using a
sinusoidal wave generator\cite{Schlick:1995:WGCG} with frequency
varying randomly about that of the runner's gait at approximately 3~Hz.
The amplitude of the additive noise was controlled by a normalized
parameter, $\lambda$, as in the torso rotation test.  A value of
$\lambda = 0$ resulted in motion data that was identical to the
original data (zero noise amplitude).  The maximum noise amplitude
used, given by $\lambda = 1$, produced a variation of $\pm0.15$~radians 
about the original joint angles
(\mbox{Figs.~\ref{fig:motPoly} and \ref{fig:motStik}}).  One
representative joint angle,  rotation of the torso at the waist about
the transverse axis, is shown in
\mbox{Figs.~\ref{fig:showRot}--C} \mbox{and~\ref{fig:dplts}--C}.

Twenty-six student volunteers who had not participated in the previous
experiments were selected as subjects.  Testing procedures were
identical to those used in the first and second experiments.


\section {Results} \label{sec-results}

To analyze the data from the experiments, we used the responses to
compute the Choice Theory sensitivity measure for each subject on each
test set.  The sensitivity measure, $log(\alpha)$, is defined as
\begin{equation}
	log(\alpha) =  { log(H/(1-H)) - log(F/(1-F))  \over  2 } ,
\end{equation}
where $H$ is the fraction of pairs in a set that were {\em different}
and which the subject labeled correctly, and $F$ is the fraction of
pairs in a section that were {\em the same} and which the subject
labeled incorrectly\cite{Mcmillan:1991:DT}.  
This measure is zero when the subject's responses are uncorrelated
with the correct responses to cause a~$50\%$~correct score, and
increases as response correlation improves, as illustrated in
Fig.~\ref{fig:svspc}.
Additionally, the measure is symmetric,
naturally invariant with respect to response bias, and suitable for
use as a distance metric\cite{Mcmillan:1991:DT}.

After sensitivity scores had been determined, a {\it post hoc}
selection criteria was used to build a subgroup of ``skilled''
subjects who had achieved a sensitivity score indicating performance
significantly better than chance with {\bf \it either} the polygonal
or the stick figure models.  Significantly better than chance was
defined as at least $73\%$ correct, which corresponds to a sensitivity
score of $log(\alpha) \geq 1.0$.  Analysis was computed both
for the group of all subjects and for the group of skilled subjects.
Sensitivity scores for each experiment averaged within subject groups
are shown in Fig.~\ref{fig:togbar}.

\figureSingleN{
  \centerline{\epsfysize=2in \epsfbox{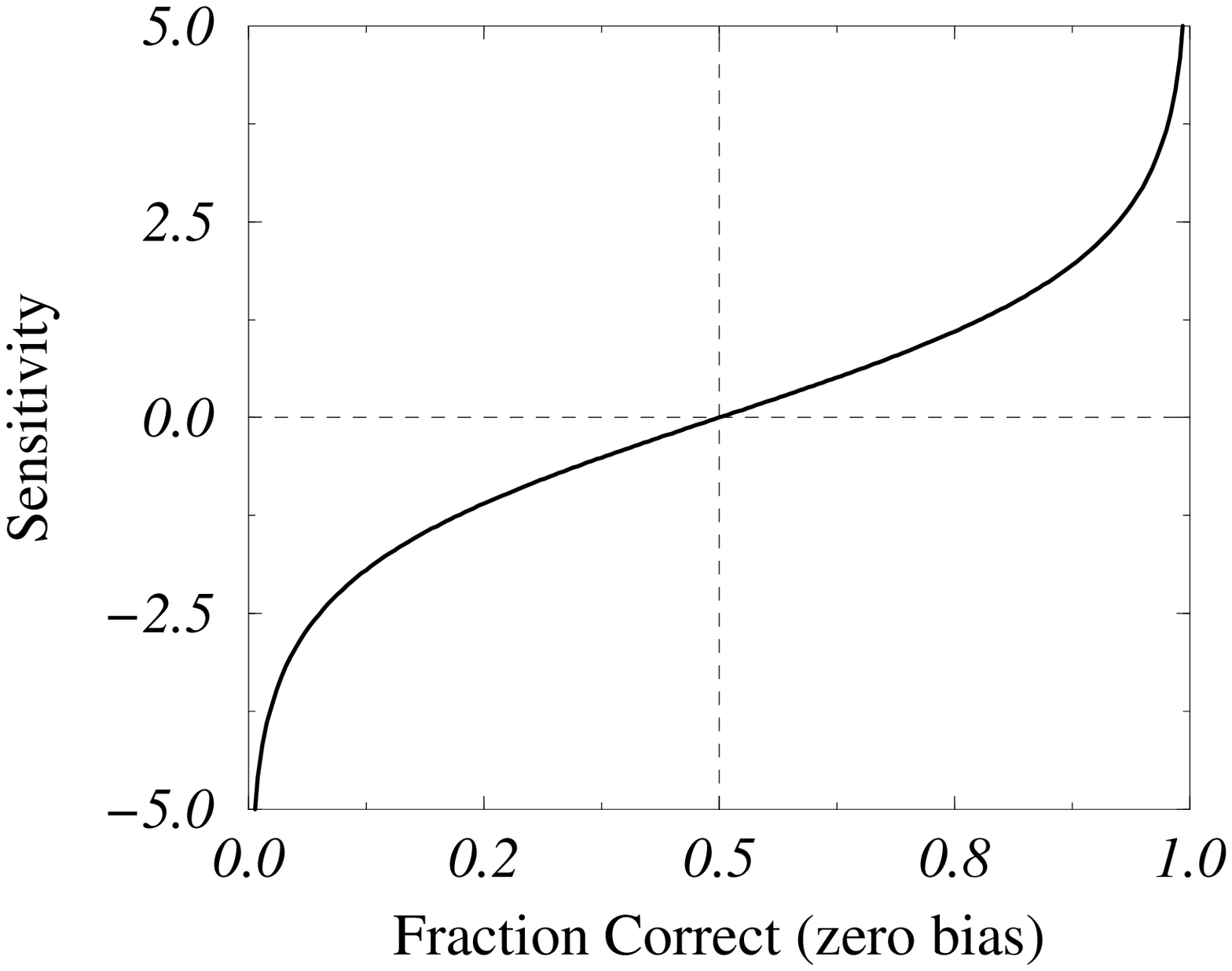}}
  \vskip -0.25in
  \caption{
    Plot of sensitivity score, $log(\alpha)$, versus fraction correct
    at zero bias.
  } \label{fig:svspc}
}

\figureSingleN{
  \centerline{\epsfxsize= 0.5 \textwidth \epsfbox{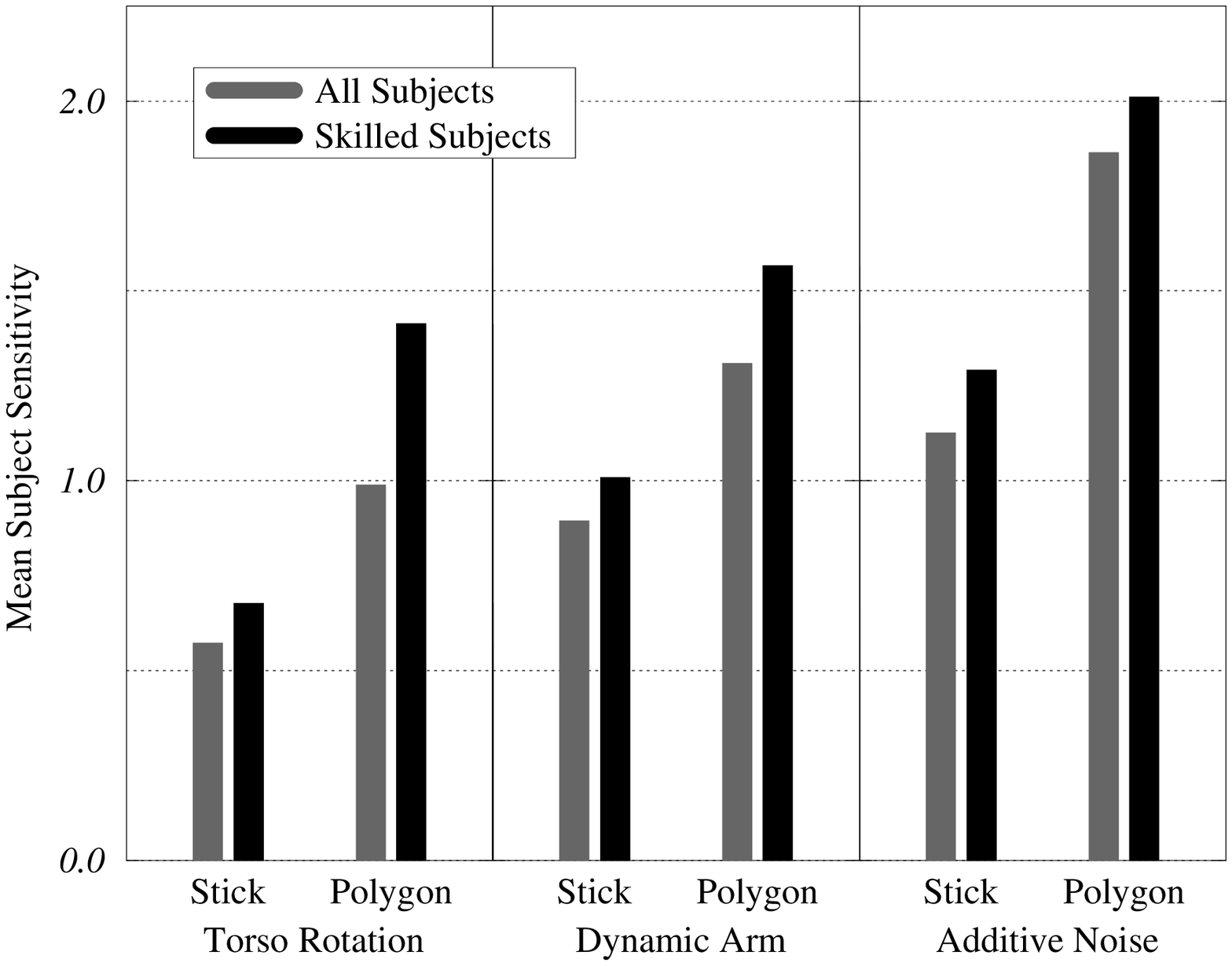}}
  \vskip -0.25in
  \caption{
    Sensitivity scores by experiment averaged over subject groups.
    Skilled subjects are those who achieved a sensitivity score of
    $log(\alpha) \geq 1.0$ on either the polygonal or the stick
    figure portion of the test.  Note that sensitivity scores are
    consistently higher with the polygonal model.
  } \label{fig:togbar}
}

In Section~\ref{sec-intro}, we proposed three possible answers to the
question of whether the geometric model used for rendering affects a
viewer's perception of motion.  The third possible answer implied that
subjects would achieve similar sensitivity measures when asked
identical questions about the motion of stick figure models or
polygonal models.  To test this hypothesis, we computed the difference
in sensitivity for each subject:
\begin{equation}
  \Delta log(\alpha) = log(\alpha_{poly})-log(\alpha_{stick}) .
\end{equation} 

The results from the three tests are summarized in
Table~\ref{tab:results}.  For the torso rotation test, the mean of the
difference in sensitivities across all subjects was $0.43$ with a
standard deviation of $0.77$.  Student's $t$--test for paired
samples\cite{Press:1992:NR} shows this difference to be significant,
$p < 0.012$.  For the group of skilled subjects, the mean rose to
$0.73$ while the standard deviation fell to $0.68$.  The $t$--test for
paired samples shows this difference to be significant, $p < 0.001$.

For the dynamic arm motion test, the mean of the difference in
sensitivities across all subjects was $0.41$ with a standard deviation
of $0.59$, a difference significant at $p < 0.003$.  For the group of
skilled subjects, the mean was $0.55$ and the standard deviation was
$0.59$, a difference significant at $p < 0.001$.

For the additive noise test, the mean of the difference in
sensitivities across all subjects was $0.74$ with a standard deviation
of $0.69$, a difference significant at $p < 0.001$.  For the group of
skilled subjects, the mean was $0.72$ and the standard deviation was
$0.73$, a difference significant at $p < 0.001$.

\begin{table*}[tb]
  \centerline{\small
    \begin{tabular}{|l||r|r|c||r|r|c|}
      \hline
      \hline			& \multicolumn{3}{c||}{All Subjects}                   	& \multicolumn{3}{c|}{Skilled Subjects}    	\\
      				& Mean 		& Std. Dev. 	& Prob. Err. 		& Mean 		& Std. Dev. 	& Prob. Err. 	\\
      \hline
      \hline Torso Rotation	& $0.43$	& $0.77$	& $p < 0.012$		& $0.73$	& $0.68$	& $p < 0.001$ 	\\
      \hline Dynamic Arm	& $0.41$	& $0.59$	& $p < 0.003$		& $0.55$	& $0.59$	& $p < 0.001$ 	\\
      \hline Additive Noise	& $0.74$	& $0.69$	& $p < 0.001$		& $0.72$	& $0.73$	& $p < 0.001$ 	\\
      \hline
      \hline
    \end{tabular}
  }
  \caption{
    Summary of results from the three experiments.  Mean and standard
    deviation are for $\Delta log(\alpha)$ by subject group.
    Probability of error is calculated with Student's $t$--test for
    paired samples.  Positive values for mean $\Delta log(\alpha)$ in
    all six test/group combinations indicate that subjects were able
    to discriminate better with the polygonal model.
  } \label{tab:results}
\end{table*}

Fig.~\ref{fig:histogramAll} shows histograms of the sensitivity
differences, $\Delta log(\alpha)$, for the three test conditions.
Positive values correspond to higher sensitivity for the set rendered
with the polygonal model.

Our results indicate that, for the three types of motion variation
tested, subjects were better able to discriminate motion variations
using the polygonal model than they were with the stick figure model.
This result holds to a high level of significance both for the
analyses computed on the group of all subjects and for the group of
skilled subjects, although the magnitude of the differences are, in
general, greater within the group of skilled subjects.

\section {Discussion} \label{sec-disc}

\figureSingleW{  
  \centerline{
    {\epsfxsize=2.1in \epsfbox{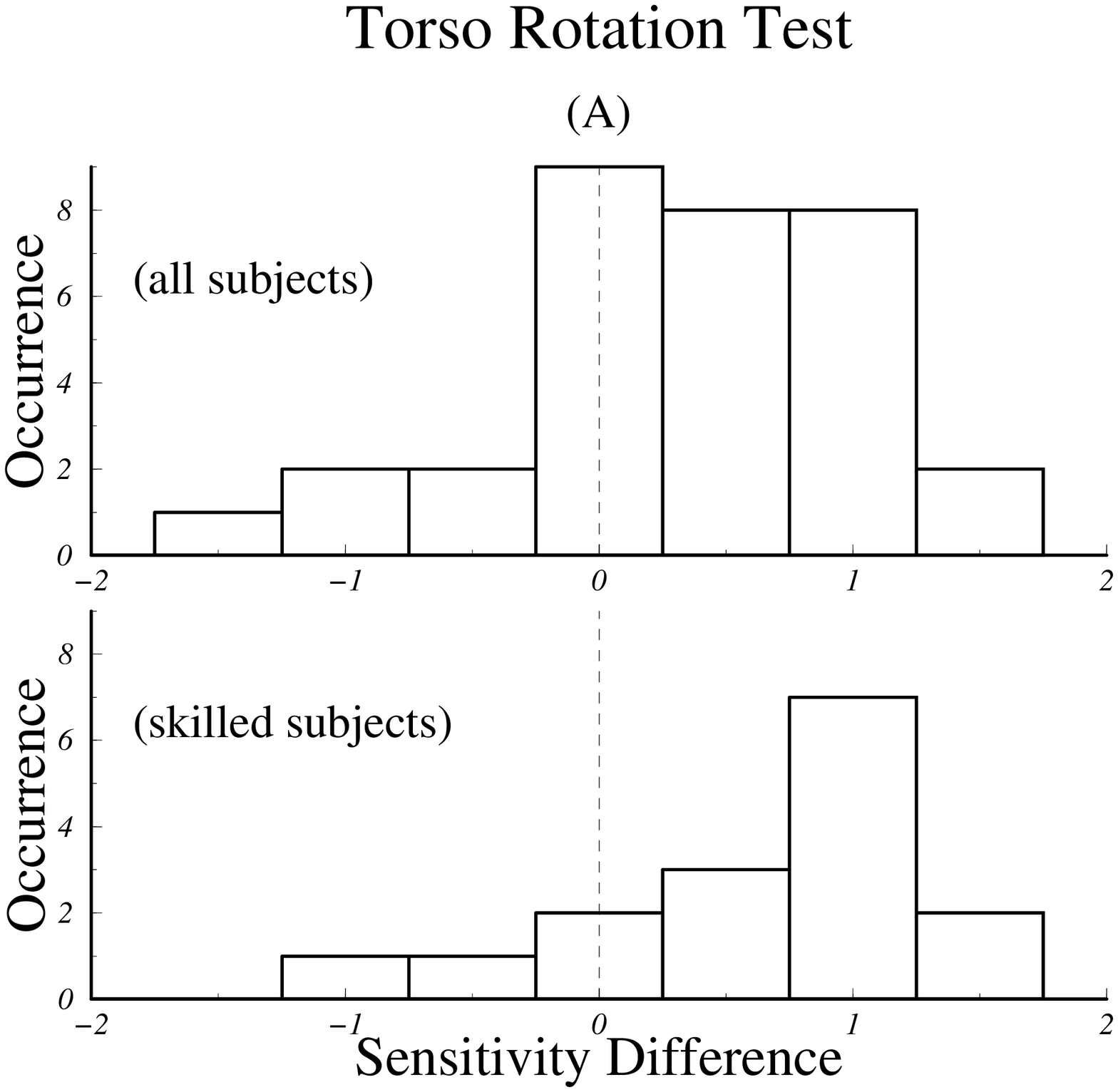}}	
    {\epsfxsize=2.1in \epsfbox{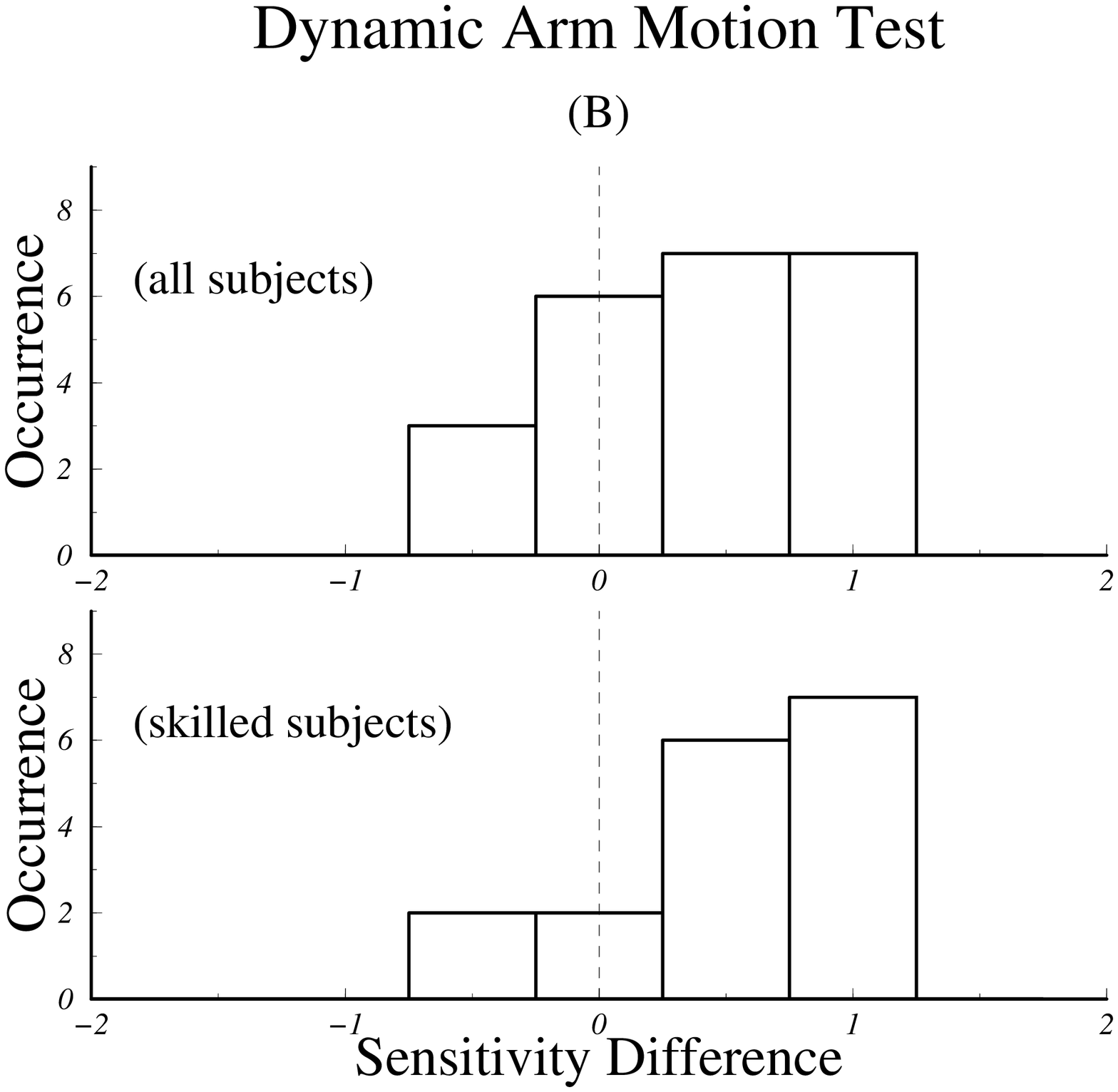}}
    {\epsfxsize=2.1in \epsfbox{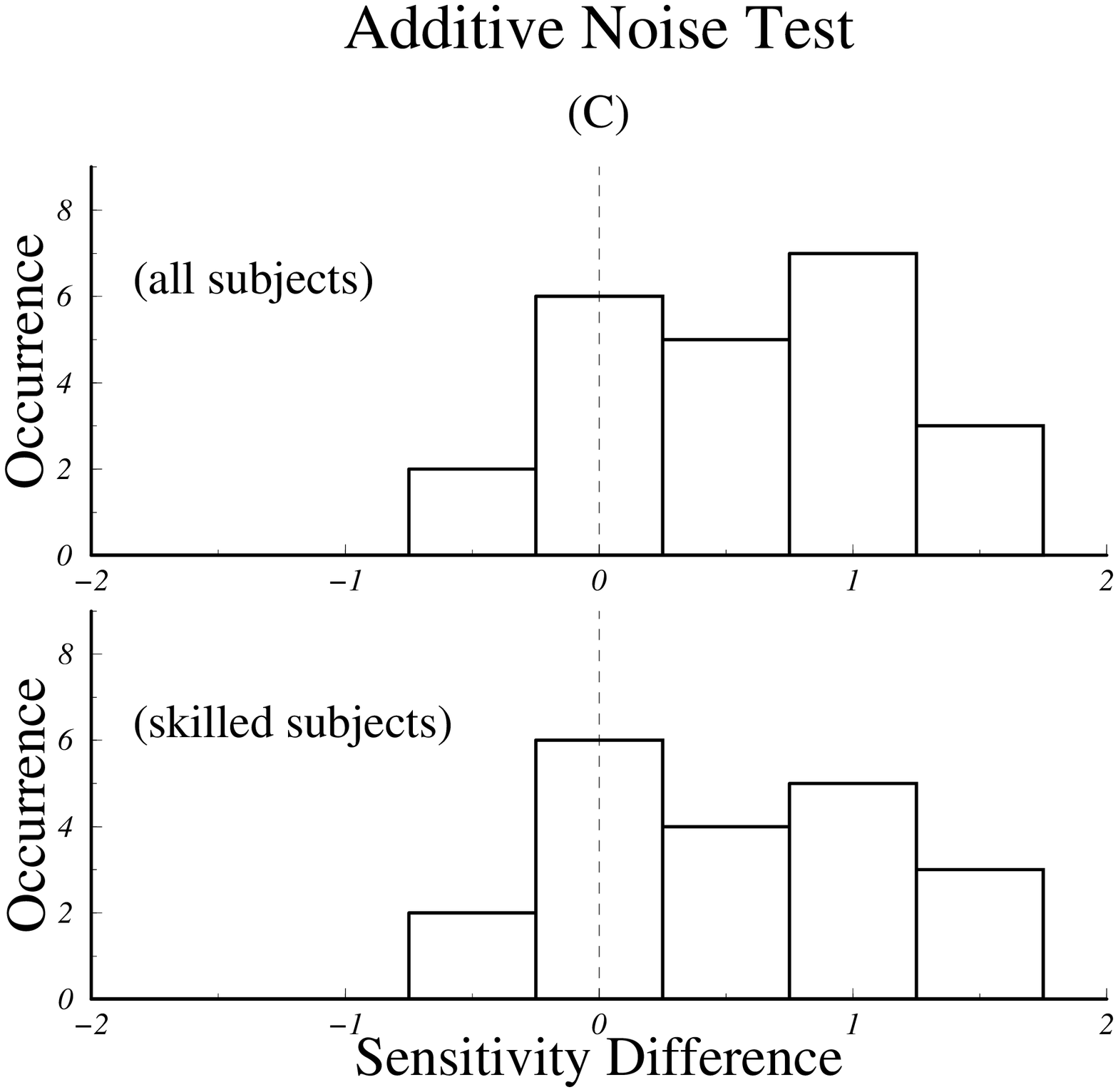}}
  }
  \caption{
    Histogram of sensitivity differences for {\bf A.} the torso
    rotation test, {\bf B.} the dynamic arm motion test, and {\bf C.}
    the additive noise test.  The upper graphs show the occurrence
    frequency for sensitivity differences, $\Delta log(\alpha)$,
    across all subjects.  The bottom graphs show the data for subjects
    who had a sensitivity of $log(\alpha) \geq 1.0$ on either
    the polygonal or the stick figure portion of the test.  Positive
    values of the sensitivity difference indicate a higher sensitivity
    to changes in the motion with the polygonal model. (Bucket size
    $=0.5$.)
  } \label{fig:histogramAll}
}

Although the differences in sensitivity measures show that our
subjects were more sensitive to motion changes when a polygonal model
was used for rendering, our results can not be generalized to
say that polygonal models are always better than stick figure
models for perceiving motions.  Rather, the two types of geometric models
are distinctly different and, in the cases we tested, polygonal models
allowed better discrimination.  There may be variations for which
the difference in sensitivity has the opposite sign, implying that
stick figures might be a better model for making fine
discriminations about that particular  motion variation.

Our results, however, do show that stick figures and polygonal models
are not equivalent for tasks that require making fine discriminations
about motion.  This observation implies that any useful comparison of
motion sequences requires that the same models and rendering methods
be used for each and might indicate that the community would benefit
from adopting a standard set of human models.  In particular,
comparing motions of a stick figure model to those of a more complex
model may be meaningless because viewer sensitivities can differ
substantially.  As a practical matter, animators may want to avoid
conducting preliminary tests only with stick figures or other simple
models because it is likely that viewers would have different
sensitivities to the more complex models that would be used in the
final rendering.

Considerable familiarity with the motion appears to make differences
in the geometric models less significant.  For example, when the
authors of this paper took the tests, they answered nearly all
questions correctly.  Of course, the authors were not included among
the subjects whose data are reported above.  If a larger subject pool
showed that subjects who were very familiar with particular animated
motions showed equal sensitivity to the two models, then we would have
evidence that using stick figures for preliminary {\it pencil tests}
of motion sequences will provide good information about the motion.
The subject, in this case the animator, is very familiar with the
motion and may be able to make subtle observations independent of the
geometric models used for rendering.

Our results do not conflict with the findings discussed in
Section~\ref{sec-background}.  Previous studies have found that
subjects were able to use a variety of models to make judgments about
human motion, these studies did not address how the subject's
proficiency might be affected.  As can be seen from
Fig.~\ref{fig:togbar}, the subjects we tested were able to make
distinctions using both polygonal and stick figure models; however
they were better able to make these distinctions when viewing motion
rendered with the polygonal model.

For the two models used in these experiments, the more complex model
was also more human-like but that may not always be the case.  Complex
but abstract models may be useful for making particular features of the
motion visible in some applications.  For example, crash test dummies
have markings for the center of mass of each body segment and other
visualization techniques such as force vectors have been 
used successfully in biomechanics research.

We used simulation combined with kinematic modifications to generate
the motion for these studies because it allowed us to control the
variations explicitly.  Motion capture data would be an interesting
source for this kind of study because it more closely matches human
motion.  However, even two consecutive captures of an actor performing
a simple task will have significant differences because of the
variability of human motion.  Capturing a set of consecutive motions
with a controlled variation for sensitivity tests would be difficult
because of this variability.

The three techniques used to modify the motion were chosen both for
their relevance to current animation techniques and for their
perceptual significance.  We chose torso rotation because previous
studies have shown that the motion of the torso provides important
cues for gender determination and subject
recognition\cite{Cutting:1978:abifgp}.  The kinematic modification
used for the torso rotation test is also similar to the modifications
an animator might make when keyframing motion or adjusting motion
capture data.  Similarly, the adjustments of the desired shoulder
joint angles used in the dynamic arm motion test are typical of the
adjustments that an animator might make to a dynamic simulation in
order to change the style of the resulting motion.  Finally, noise is
found in naturally occurring motions, and additive noise
generators have been used to synthesize natural and appealing human
motion\cite{Perlin:1995:RTR}.

A potential problem with the experimental design used in this study is
that the test must be of an appropriate difficulty.  If the test is
too difficult, then subject responses will be guesses
regardless of which model is presented.  Conversely, if the test is
too easy, then all subject responses will be correct.  In
either case, the data gathered will not be useful.  We can increase or
decrease the difficulty of a test by changing the spacing of the
$\lambda$ values for the trials or the amount of information given to
the subjects about the alterations to the motion.  Unfortunately, it
can be difficult to devise a test sequence of appropriate difficulty.
This problem could be overcome by using tests that adaptively adjust
difficulty level by selecting subsequent questions based on past
responses.  Alternatively, selection criteria can be used to cull
subjects whose responses are not significantly correlated with the
test stimuli.

While our assessment that the polygonal models allow greater
sensitivity holds irrespective of culling, it is interesting to note
how selection based on performance criteria does affect the data.  As
can be seen from the average scores shown in Fig.~\ref{fig:togbar},
subjects who took the torso rotation test achieved lower scores than
did those who took the additive noise test, probably because the torso
rotation test was more difficult.  Comparing the results of the torso
rotation test before and after culling shows that the mean of $\Delta
log(\alpha)$ as well as the shape of the histograms in
Fig.~\ref{fig:histogramAll}.A were notably different between the
group of all subjects and the group of skilled subjects.  For the
easier, additive noise test, the selection criteria has essentially no
effect.  Moreover, the effect of the selection criteria on the torso
rotation data appears to make it more closely resemble the data from
the additive noise test, thereby supporting the notion that lowering
the difficulty of the test and selecting subjects based on performance
criteria are approximately equivalent.

Although we did not formally measure the subjects' perceptions of how
well they did on the test, it appeared that their perceptions did not
always match their performance.  Several subjects were certain that
they had scored higher on the section with the stick figure model when
in fact they had a higher sensitivity to motion changes with the
polygonal model.

To create the animation sequences for these tests, we altered only the
motion and the geometric models used; all other aspects of the
rendering were held constant.  It would be interesting to explore
whether, and how, other aspects of the rendering affect the perception
of motion as well as whether these results hold for behaviors other than
running.  For example, we have informally observed that the motion
of the simulated runner appears more natural when the tracking camera
has a constant velocity rather than one that matches the periodic
accelerations of the runner's center of mass.  When the camera motion
matches the acceleration of the center of mass exactly, the running
motion appears jerky.  More sophisticated models that incorporate
clothing and skin may help to smooth out rapid accelerations of the
limbs and make the motion appear more natural.  Motion blur probably
plays a similar role.  Textured ground planes and shadows help to
determine motion of the feet with respect to the ground and may
provide important clues about the details of the motion.

If we had enough psychophysical results to build a model of how people
perceive motion, we could optimize the rendering of animated sequences
by emphasizing those factors that would make the greatest differences
in how a viewer perceives the sequence either consciously or
unconsciously.  This approach of using
results from the psychophysical literature to refine rendering
techniques has already been used successfully for still
images\cite{Ferwerda:1996:MVARIS,Kawai:1993:RGBR,Teo:1994:PID}.


\section*{Acknowledgments}  \label{sec-ack}

The authors would like to thank Jacquelyn Gray, John Pani, Neff
Walker, and the reviewers for their valuable comments.  This project
was supported in part by NSF NYI Grant No. IRI-9457621, Mitsubishi
Electric Research Laboratory, and a Packard Fellowship.  An earlier
version of this work, reporting preliminary results, appeared in {\it
The Conference Proceedings of Graphics Interface '97}.

\bibliography{rencmp}

\begin{biographyPIC}{Jessica K. Hodgins}{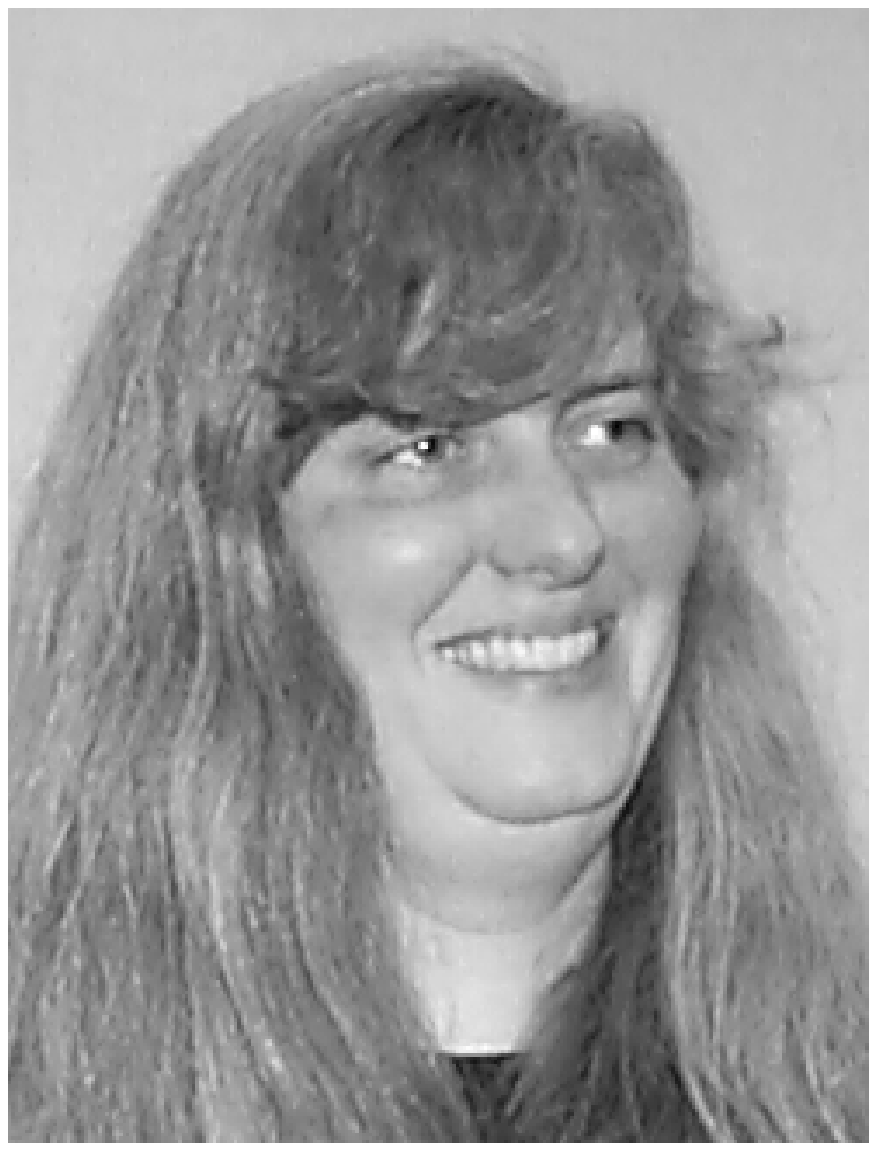}
received her Ph.D. from the Computer Science Department at Carnegie
Mellon University in 1989.  From 1989 to 1992, Hodgins was a
postdoctoral researcher in the MIT Artificial Intelligence Laboratory
and the IBM Thomas J. Watson Research Center.  She is currently an
associate professor in the College of Computing at the Georgia
Institute of Technology and a member of the Graphics, Visualization
and Usability Center.  Her research explores techniques that may
someday allow robots and animated creatures to plan and control their
actions in complex and unpredictable environments.  In 1994 she
received an NSF Young Investigator Award and was awarded a Packard
Fellowship.  In 1995 she received a Sloan Foundation Fellowship.  She
is on the Editorial Board of the Journal of Autonomous Robots, the
IEEE Magazine on Robotics and Automation and ACM Transactions on
Graphics.
\end{biographyPIC}

\begin{biographyPIC}{James F. O'Brien}{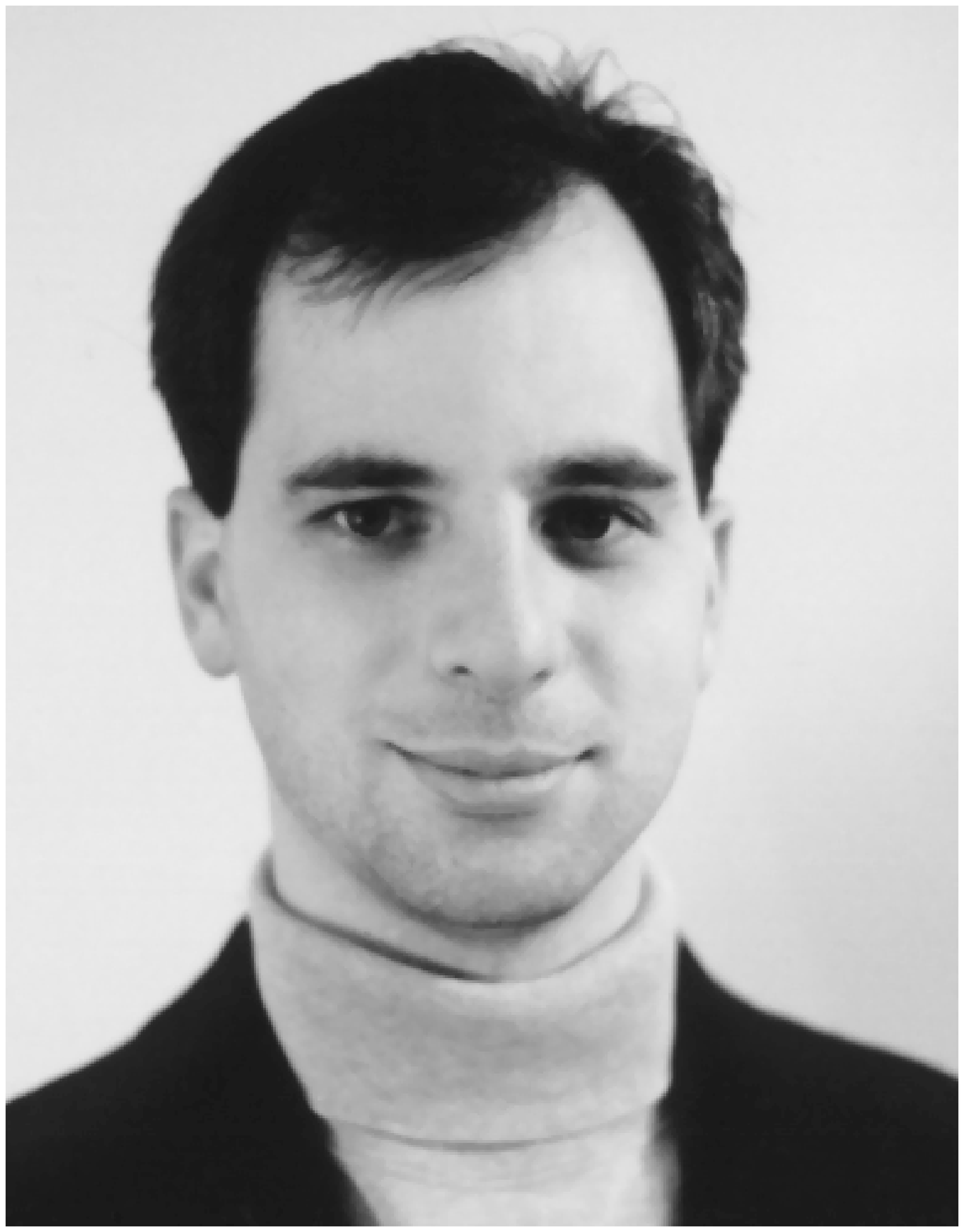}
is a doctoral student in the College of Computing at the Georgia
Institute of Technology, and a member of the Graphics, Visualization,
and Usability Center.  He received a M.S. in Computer Science from the
Georgia Institute of Technology in 1996 and a B.S. in Computer Science
from Florida International University in 1992.  His research interests
include physically based animation and geometric modeling.  In 1997 he
received a graduate fellowship from the Intel Foundation.
\end{biographyPIC}

\begin{biographyPIC}{Jack Tumblin}{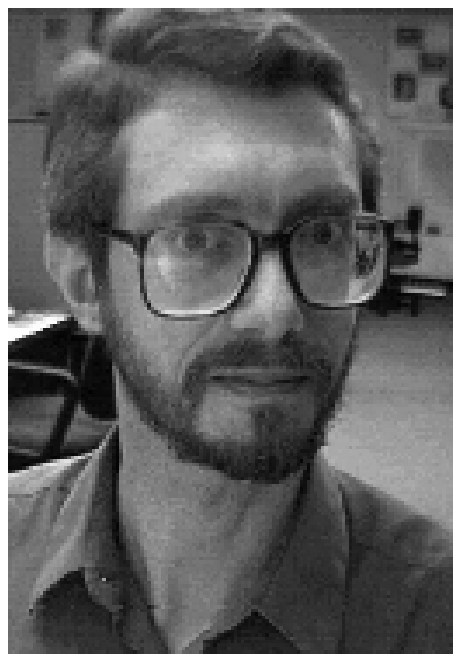}
is a PhD candidate in computer science.  His research interests
include computer graphics, visual perception and image processing.
After graduating from Georgia Tech with a BSEE in 1979, he worked as a
television broadcast engineer and later designed flight simulator
image-generating hardware, resulting in four patents. He returned to
Georgia Tech and in 1990 earned an MS (EE) degree.
\end{biographyPIC}


\vfill\eject
\end{document}